\documentclass[twocolumn,pra,tighten,floatfix,showpacs]{revtex4}
\usepackage{graphicx}

\def\<{\langle}
\def\>{\rangle}

\def\be{\begin{equation}}
\def\ee{\end{equation}}

\begin{document}
\preprint{cond-mat} \title{Full counting statistics in a disordered free fermion system}

\author{G. C. Levine,  M. J. Bantegui and J. A. Burg}

\address{Department of Physics and Astronomy, Hofstra University,
Hempstead, NY 11549}

\date{\today}

\begin{abstract}
The Full Counting Statistics (FCS) is studied for a one-dimensional system of non-interacting fermions with and without disorder. For two unbiased $L$ site lattices connected at time $t=0$, the charge variance increases as the natural logarithm of $t$, following the universal expression $\langle \delta N^2\rangle \approx  \frac{1}{\pi^2}\log{t}$. Since the static charge variance for a length $l$ region is given by $\langle \delta N^2\rangle \approx  \frac{1}{\pi^2}\log{l}$,  this result reflects the underlying relativistic or conformal invariance and dynamical exponent $z=1$ of the disorder-free lattice. With disorder and strongly localized fermions, we have compared our results to a model with a dynamical exponent $z \ne 1$, and also a model for entanglement entropy based upon dynamical scaling at the Infinite Disorder Fixed Point (IDFP).  The latter scaling, which predicts $\langle \delta N^2\rangle \propto \log\log{t}$, appears to better describe the charge variance of disordered 1-d fermions. When a bias voltage is introduced, the behavior changes dramatically and the charge and variance become proportional to $(\log{t})^{1/\psi}$ and $\log{t}$, respectively. The exponent $\psi$ may be related to the critical exponent characterizing spatial/energy fluctuations at the IDFP. 
\end{abstract}

\maketitle
\section{Introduction} 

Full Counting Statistics (FCS) in fermion systems \cite{Levitov_Lesovik} have attracted attention recently for their connection with entanglement entropy \cite{KL_noise,lehur}. Specifically, it has been shown that the von Neumann entanglement entropy associated with the coupling between two conductors may be expressed as a series involving the cumulants of conserved charge in one of the conductors \cite{KL_noise,lehur}.  In some sense, noise in an unbiased junction between two conductors may be thought of as the propagation of entanglement after the junction is closed. In a gapless, disorder-free fermionic system, it appears that the rate of propagation is effectively the Fermi velocity, $v_F$.  It is natural, then, to ask if entanglement propagates when there is strong disorder and the fermions are completely localized?

Consider a situation where the conductors are 1-dimensional and the barrier between them  reflectionless. In the unbiased case, the second cumulant of charge and entanglement entropy are simply proportional (all higher cumulants vanish). For two 1-dimensional noninteracting fermion systems of length $L$, connected at time $t=0$, the charge variance increases logarithmically in $t$, following the universal expression $\langle \delta N^2\rangle \approx  \frac{1}{\pi^2}\log{t}$, for $t$ much shorter than the ballistic time to encounter the boundary, $t_{b} \sim L$ \cite{Levitov_Lesovik}. However, the equilibrium charge variance, at zero temperature, for a length $l$ region where $l<<L$ is given by a similar result: $\langle \delta N^2\rangle \approx  \frac{1}{\pi^2}\log{l}$.  These results may be roughly interpreted in the following way: after time, $t$, charge fluctuations only reflect the participation of those fermions within a distance $l = v_F t$ of the junction \cite{cardy_quench}. Entanglement entropy, $S$, of the half-chain exhibits the same scaling in space and time but with a different universal prefactor: $S = \frac{\pi^2}{6} \langle \delta N^2\rangle$ \cite{KL_noise}.   Whether calculating noise or entropy, these result reflects the underlying relativistic or conformal invariance and dynamical exponent $z=1$. 

Although the calculations of entropy and noise are quite different, both results express underlying universality unique to 1-dimensional (critical) systems.  It is also well known that some features of universality extend to entanglement entropy computed in 1-dimensional disordered systems. Disorder is strongly relevant in 1-dimension and in the case of noninteracting fermions it results in complete localization, described by the Infinite Disorder Fixed Point (IDFP) ) \cite{RSP}. Remarkably, entanglement entropy retains its universal logarithmic form but with a modified central charge $c_{\rm eff} = c\log{2}$ \cite{Refael_Moore}.  Similarly the equilibrium charge variance of a size $l$ region is given by $\langle \delta N^2\rangle \approx  \frac{\log{2}}{\pi^2}\log{l}$.

Since even fully localized fermions exhibit $\log{l}$ entropy and noise in equilibrium---similar to delocalized ones---one might expect that given enough time the system would relax to a state resembling the IDFP with long range entanglements straddling the junction after it is closed. If nothing else, the fermion wavefunction must maintain antisymmetry as fermions cross the junction, which is generally a sufficient condition for entanglement.  

In this manuscript we computationally explore the temporal development charge fluctuations when strong disorder is present. Questions similar to these have been addressed in an interesting analytical study of Lieb-Robinson type bounds in spatially disordered systems \cite{light_cone} and in computational works on entanglement after a quench \cite{Igloi_disorder,dis_dynamics}.  Very recently, the growth of entanglement after a quench has been studied in connection with the role of interactions \cite{moore_disorder}.  We also note that for strictly 1-dimensional conductors, disorder is always relevant and one would expect the long time limit of FCS to be controlled by IDFP physics. For this reason a numerical analysis of FCS in a disordered 1-dimensional lattice might be relevant for experiments on noise in quasi 1-dimensional conductors.  These results may also be relevant to recent experiments on cold atom gases in random lattices \cite{CAG_disorder}.

\section{The model}

Consider two one-dimensional noninteracting systems of spinless fermions (subsystems $A$ and $B$), connected by a weak link at one site (fig. 1). The model hamiltonian is:
\begin{eqnarray}
\label{ham}
H &=& -\sum_{\< x,y \> \alpha=A,B}{\kappa_{xy}(c^{\alpha \dagger}_{x}c^\alpha_{y} + c^{\alpha \dagger}_{y}c^\alpha_{x}})\\
\nonumber&-& w (c^{A \dagger}_{0}c^B_{0} + c^{B \dagger}_{0}c^A_{0})\\
&=& H_0 + H_w
\end{eqnarray}
where $x$ and $y$ are one-dimensional site indices and $A$ and $B$ denote the two systems with nearest-neighbor hopping amplitudes $\kappa_{xy}$ and weak link amplitude, $w$.  Each subsystem consists of $L$ sites and is taken to have fixed boundary conditions at its ends. Furthermore, we will restrict our considerations to the case of $L$ fermions in $2L$ sites. $c_{x}^{\alpha}$ ($c^{\alpha \dagger}_x,$) destroys (creates) fermions at site $x$ in subsytem $\alpha$ and obeys the conventional fermion algebra. 
\begin{figure}[ht]
\includegraphics[width=8.0cm]{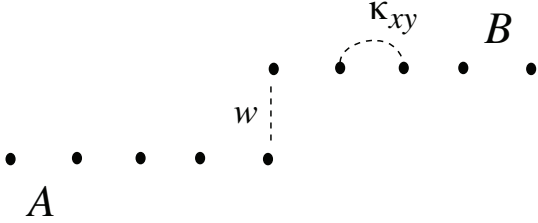}
\caption{\label{fig2} Two fermion lattices with random hopping amplitudes $\kappa_{xy}$, coupled by a weak link with amplitude $w$.}
\end{figure}

Without disorder, all $\kappa_{xy}$ will be set to $\kappa = 1$. When disorder is included in our calculation, the set $\{\kappa_{xy}\}$ will be chosen from a uniform random distribution over the interval $[\kappa - \delta \kappa, \kappa + \delta \kappa]$. If the $\{\kappa_{xy}\}$ are chosen independently for the two subsystems, $A$ and $B$ will be at different chemical potentials. Given exactly $L/2$ fermions occupying each subsystem initially, there will be a net particle flux in some direction as the system evolves away from its initial condition.  To study solely the fluctuation in the number of particles in one subsystem, we adopt identical realizations of disorder in $A$ and $B$ subsystems. Although unphysical, this simplifies the analysis of the fluctuations.

At time $t<0$, the two subsystems $A$ and $B$ are disconnected and prepared in the ground state of noninteracting fermions:
\begin{equation}
|\psi\rangle = \prod_{k=1}^{L/2}{c_k^{A\dagger}c_k^{B\dagger}|0\rangle}
\end{equation}
where (without disorder) the operator $c^\dagger_k $ is related to $c^\dagger_x$ in the usual way:
\begin{equation}
\label{unperturbed}
c^\dagger_x = \sqrt{\frac{2}{L+1}} \sum_{k=1}^L{\sin{\frac{k\pi x}{L+1}} c^\dagger_k} = \sum_{k=1}^L{\phi_k(x)
c^\dagger_k} 
\end{equation}
With disorder, $\phi_k(x)$ will be an eigenfunction of the $L \times L$ matrix $\kappa_{xy}$:
\begin{equation}
\sum_{y=1}^L\kappa_{xy} \phi_k(y) = \epsilon_k \phi_k(x)
\end{equation}
It will be useful later to rewrite $H$ in the momentum basis
\begin{eqnarray}
\label{mom_ham}
H &=& \sum_{k=1}^L{\epsilon_k(c^{A \dagger}_{k}c^A_{k} + c^{B\dagger}_{k}c^B_{k}})\\
\nonumber&+& \sum_{k,k^\prime = 1}^L{M_{k k^\prime} (c^{A\dagger}_k c^B_{k^\prime} + c^{B\dagger}_k c^A_{k^\prime} )}\\
&=& \sum_{kk^\prime=1; \alpha\alpha^\prime=A,B}^L{h_{kk^\prime}^{\alpha\alpha^\prime}c^{\alpha\dagger}_k c^{\alpha^\prime}_{k^\prime}}
\end{eqnarray}
where $M_{kk^\prime} \equiv w \phi_k(1)\phi_{k^\prime}(1)$. In the last equality, $h_{kk^\prime}^{\alpha\alpha^\prime}$ may be thought of as unfolded into a $2L$-dimension single particle hamiltonian matrix.

At time $t=0$, the subsystems are connected with amplitude, $w$.  The probability, $b_N(t)$, that $N$ particles are found in system $A$ at time $t$ is commonly expressed in terms of a generating function:
\begin{equation}
\chi(\lambda,t) \equiv \sum_{N=1}^{L}{b_N(t)e^{i \lambda N}}
\end{equation}
Following Klich \cite{Klich_trace}, the generating function $\chi(\lambda,t )$ corresponds to the following quantum expectation value,
\begin{eqnarray}
\chi(\lambda,t) &=& \langle \psi| U^\dagger e^{i \lambda \sum_{k=1}^L{c_k^{A\dagger}c_k^A}} U |\psi\rangle\\
&=&  \langle \psi| U^\dagger e^{i \lambda \sum_{\alpha\alpha^\prime kk^\prime}{P_{kk^\prime}^{\alpha\alpha^\prime}c_k^{\alpha\dagger}c_{k^\prime}^{\alpha^\prime}}} U |\psi\rangle\\
&=&  \langle \psi| e^{i \lambda \sum_{\alpha\alpha^\prime kk^\prime}{P_{kk^\prime}^{\alpha\alpha^\prime}(t)c_k^{\alpha\dagger}c_{k^\prime}^{\alpha^\prime}}}  |\psi\rangle
\end{eqnarray}
where $U \equiv e^{-i H t}$ is the full many body evolution operator. In the second equality, $P_{kk^\prime}^{\alpha\alpha^\prime}$ is the $2L$-dimension projection operator, written in momentum space, onto the $A$ subsystem.  $\chi(\lambda,t )$ is then expressed (the third equality) in terms of the time evolution of the projection operator in the single particle Hilbert space (dropping indices for clarity): $P(t) = e^{i h t} P e^{-i h t}$.

Now the expectation value with respect to $|\psi\rangle$ is replaced with a trace over an initial density matrix, $\rho = \frac{1}{Z}e^{-\beta H_0}$, corresponding to two independent Fermi systems ($A$ and $B$), at inverse temperature, $\beta$, where $Z \equiv {\rm tr }{e^{-\beta H_0}} $. The Klich trace formula is then applied in the zero temperature limit yielding,
\begin{eqnarray}
\chi(\lambda,t) &=&\lim_{\beta \rightarrow \infty} \frac{1}{Z}{\rm tr } (e^{-\beta H_0}e^{i\lambda P(t)})\\
&=& {\rm det}[1 - N + Ne^{i \lambda P(t)} ]
\end{eqnarray}
where $N$ is the projector onto the Fermi sea of the two subsystems; in matrix form, $N = \delta_{\alpha \alpha^\prime} \delta_{k k^\prime} \theta(k-L/2)$.

Using the property $P^2(t) = P(t)$, $\chi(\lambda,t)$ may be expressed in terms of the $L$ nonzero eigenvalues of $NP(t)$ denoted by $\{ p_m(t)\}$. 
\begin{equation}
\label{fcs}
\chi(\lambda,t) = \prod_{m=1}^{L}{(1-p_m(t) + p_m(t)e^{i\lambda})}
\end{equation}
The probabilities $\{b_N(t)\}$ may be computed from eq. (\ref{fcs}) using a clever recursive scheme described by Schonhammer in \cite{Schonhammer_num1,Schonhammer_num2}.

\begin{figure}[ht]
\includegraphics[width=7.5cm]{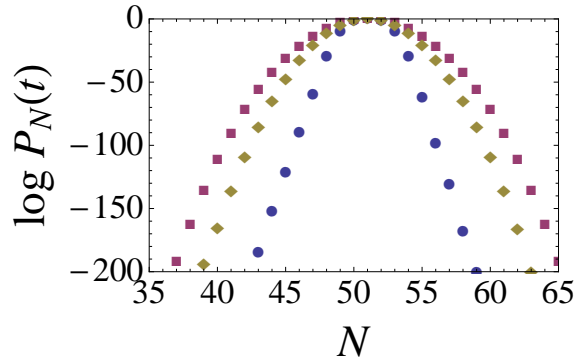}
\caption{\label{fig2} (Color online) Natural logarithm of the probability distribution $b_N(t)$ for times $t = 1, 9, 18$ ($\circ, \diamond, \sqcap$, respectively).  For $t=18$, the probability distribution is very close to a Gaussian distribution in agreement with the known behavior for a reflectionless barrier as reported in \cite{Levitov_Lesovik}. }
\end{figure}

\section{results for full counting statistics without disorder}

In the $L \rightarrow \infty$, continuum limit of the disorder free model described above exact results are known for the FCS generating function. When the junction between $A$ and $B$ is reflection-less and unbiased, the generating function is exactly Gaussian with a variance that depends logarithmically on time \cite{Levitov_Lesovik}:
\begin{equation}
\label{gaussian_fcs}
\chi(\lambda,t) = e^{-\frac{\lambda^2}{2\pi^2}\log{\kappa t}}
\end{equation}
Since the probability distribution $b_N(t)$ is related to $\chi(\lambda,t)$ by Fourier transform, $b_N(t)$ follows a Gaussian distribution with respect to $N$ for all times. Fig. 2 shows results for $\log{b_N(t)}$ computed on an $L=100$ site, ordered lattice.  For the representative times shown, the distribution is Gaussian with a variance that  increases in time.  Noting the vertical scale of the graph, the variances all involve fluctuations of $O(1)$ fermions. 
\begin{figure}[ht]
\includegraphics[width=7.5cm]{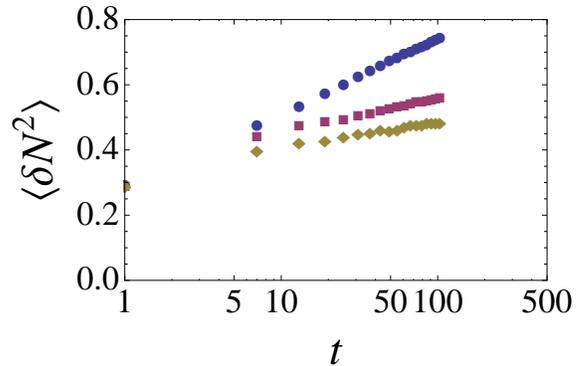}
 \caption{\label{fig2} (Color online) Variance as a function of time, $t$, for an $L=600$ lattice. Calculation for variance in translation invariant lattice ($\circ$) exhibits universal $\frac{1}{\pi^2}\log{t}$ behavior. Variance computed for disordered lattices with disorder strength $\delta \kappa = 0.3$ ($\sqcap$) and $\delta \kappa = 0.5$ ($\diamond$) shown for comparison. }
\end{figure}

To examine the Gaussian probability distribution given by eq. (\ref{gaussian_fcs}), we calculate the variance,
\be
\langle \delta N^2(t) \rangle \equiv \sum_{N=1}^L{N^2 b_N(t)} - (\sum_{N=1}^L{N b_N(t)})^2
\ee
For a system without disorder, we first look in the regime $1<< t << \kappa L$ where the ballistic propagation of fermions to the boundary is irrelevant. The topmost data in fig. 3 shows the time development of the variance in an $L=600$ site system for $t \le 100$. A least squares fit yields 
\be
\label{logt}
\langle \delta N^2(t) \rangle \approx 0.282 + 0.0985 \log{t}
\ee
agreeing within less than $3\%$ of the universal value for the prefactor of the logarithmic term: $1/\pi^2 \approx 0.1013$. 
\begin{figure}[ht]
\includegraphics[width=7.5cm]{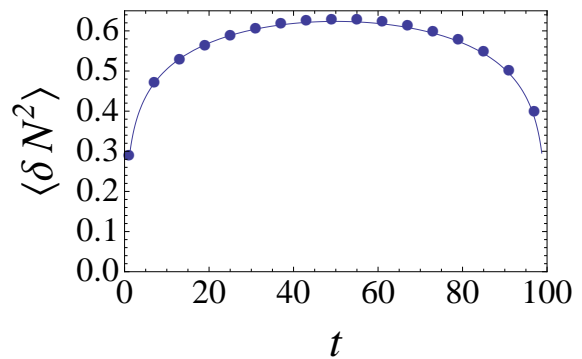}
\caption{\label{fig2}  Variance as a function of time for an $L =100$ site lattice showing features of conformal invariance.  Solid line is graph of equation (\ref{FCS_CFT}).}
\end{figure}

When the variance is computed in the regime $t\in[0,  \kappa L]$ (fig. 4), where the counting statistics are affected by the boundaries, the variance reaches a maximum at the subsystem half-size, $L/2$. In terms of the relationship between noise and entanglement entropy, this result might be anticipated from the CFT expression for entropy after a local quench \cite{stephan_quench}.  Klich and Levitov \cite{KL_noise} showed that for a reflectionless barrier, entanglement entropy at time $t$ after the channel is closed is given simply by the second cumulant of the FCS for same geometry.    Fig. 4 shows the variance computed for an $L = 100$ lattice; the data is well fit by an expression,
\be
\label{FCS_CFT}
\langle \delta N^2(t) \rangle \equiv 0.273 + \frac{1}{\pi^2}\log{|\frac{L}{\pi}\sin{\frac{\pi t}{L}}|}
\ee
This result agrees with the analytical time dependence for entanglement entropy after a local quench in \cite{stephan_quench} and earlier numerical results that anticipated the CFT result (second reference of \cite{stephan_quench}). We also note that this expression may be identified with equation (21) of reference \cite{KL_noise} by replacing the the spatial interval $L$ with the time window, $T$, as defined in the reference.

\section{equilibrium charge fluctuations}

To help frame our results about FCS with disordered fermions we will first discuss equilibrium charge fluctuations. It has been noted that the logarithmic behavior of the variance, illustrated in eq. (\ref{logt}), is a reflection of the underlying 1-d nature (in time) of the FCS problem in any dimension \cite{Levitov_Lesovik}.  Due to relativistic invariance, the time scale, $t$, sets a maximum length scale, $l = v_F t$ outside of which number fluctuations are uncoupled.  Since number fluctuations in a length $l$ segment of a 1-d channel are proportional to $\frac{1}{\pi^2}\log{l}$, the charge fluctuations in a channel closed for a period of time $t$ cannot exceed $\frac{1}{\pi^2}\log{t} + c^\prime$, where $c^\prime$ is a nonuniversal constant.  Equation (\ref{FCS_CFT}) is also a reflection of the same space/time symmetry through the stronger constraint of conformal invariance.

\begin{figure}[ht]
\includegraphics[width=7.5cm]{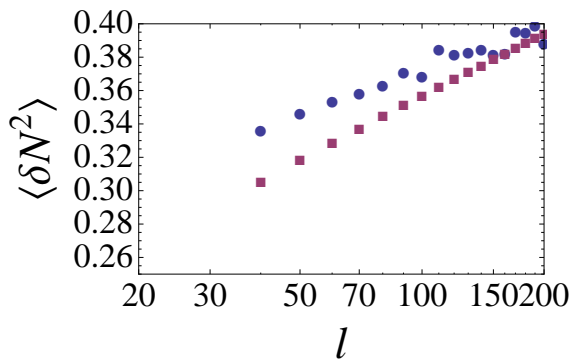}
\caption{\label{fig2}  Equilibrium charge fluctuations for a disordered half-chain of length $l$ with disorder strength $\delta \kappa = 1.0$ averaged over 2000 realizations ($\circ$) and disorder-free half chain ($\sqcap$) . }
\end{figure}

Although it is a well known result, in Figure 5 (lower data set) we compute the charge variance of the $A$ subsystem in an $L$-site disorder free system. The equilibrium charge fluctuations are computed from the fermion correlation function, $g(xy) = \langle c(x) c^\dagger(y)\rangle$ evaluated in the ground state of $L/2$ fermions in an $L$ site system:
\be
\langle N^2\rangle - \langle N\rangle^2 = \frac{l}{2} - \sum_{x,y\in C}{g(xy)^2}
\ee
A least squares fit of the data yields the dependence on $l$, the number of sites in subsystem $A$,
\be
\label{spatial_fluct}
\langle \delta N^2 \rangle = a + \frac{b}{\pi^2} \log{l} 
\ee
close to the universal result $\frac{1}{2\pi^2}\log{l}$  ($a \approx 0.10, b \approx 0.53$). The factor of $1/2$ appears in front of the logarithm because there is only a single boundary.  

For one dimensional noninteracting fermions, disorder is a relevant perturbation and leads to the IDFP corresponding to localized fermions. It is known that entanglement entropy of a conformally invariant system becomes modified with disorder, the entropy following the same logarithmic behavior but with a modified central charge $c_{\rm eff} = c\log{2}$ \cite{Refael_Moore}.  It was shown in \cite{refael_numberfluct} that the variance of the $z$-component of spin in the Random Singlet Phase (RSP)\cite{refael_numberfluct} is proportional to the entanglement entropy; based on this result, the charge variance for fermions follows equation (\ref{spatial_fluct}) with $b=\frac{1}{2}\log{2} \approx 0.347$. Our small disordered lattice calculation, averaging over $2000$ realizations of disorder (upper data set of figure 5), agrees reasonably well with the known result yielding $a \approx 0.20$ and $b \approx 0.365$. Since charge fluctuations at the fixed point should be independent of the microscopic disorder strength, we used a strong disorder parameter $\delta \kappa=1.0$; presumably our close agreement reflects that we are close to the fixed point with this choice.

\section{results for FCS with disorder}

The IDFP corresponds to completely localized fermions.  Therefore, one would expect that the heuristic argument for the $\frac{1}{\pi^2}\log{t}$ dependence of the variance---specifically, that FCS corresponds to the number fluctuations in a $L = v_F t$ size spatial window---should fail.  Since the gate time $t$ sets an effective ballistic length scale $L = v_F t$ for fermions in a translation invariant lattice, for weak disorder one might expect a ballistic regime wherein the variance behaves as $\frac{1}{\pi^2}\log{t}$ for $v_F t < \xi$ where $\xi$ is the elastic mean free path or localization length.  Beyond this ballistic regime, at $v_F t > \xi$, the variance would saturate to a value proportional to $\log{\xi}$.

For a system with disorder, we first look in the regime $1<< t << \kappa L$ where propagation of fermions to the boundary is believed to be irrelevant based upon our prior results without disorder. The two lower data sets in fig. 3 shows the time development of the variance in an $L=600$ lattice for two different disorder strengths, $\delta \kappa = 0.3, 0.5$.  Clearly there is a large reduction in the variance and a deviation from the universal result for the disorder-free system (upper data set). 

The results of a disorder FCS calculation for an $L=600$ lattice with $\delta \kappa = 1.0$) extended to $t=10^4$ is shown in figure 6.  We will discuss the data by fitting it to an expression of the form:
\be
\label{variance_disorder}
\langle \delta N^2(t) \rangle = a + \frac{b}{\pi^2} \log{t}
\ee
with $b=1$ corresponding to the variance without disorder. ($\kappa=1$ has been suppressed in the argument of the logarithm.)  

In the presence of strong disorder, the charge variance appears to be logarithmic in time, but with a prefactor significantly smaller than that of the disorder-free result. A least squares fit to the data of figure 6 gives $a=0.27, b=0.138$.

\begin{figure}[ht]
\includegraphics[width=7.5cm]{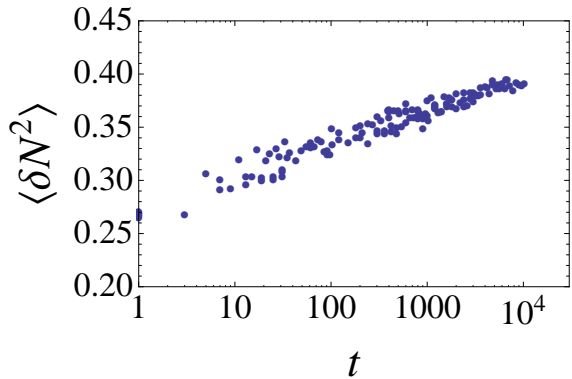}
\caption{\label{fig2} FCS computation of charge variance for a disordered $L=600$ lattice with $\delta \kappa = 1.0$. Results were averaged over $1000$ realizations of disorder.}
\end{figure}

Figure 7 shows a comparison of the time dependence of the variance for three disorder strengths, $\delta \kappa = 0.5,0.75$ and $1.0$.  Although it is difficult to make a definitive interpretation, for weak disorder there appears to be a ballistic regime for $t < 10$.  At longer times the variance for the weakest disorder strength, $\delta \kappa = 0.5$, seems to show logarithmic behavior with an initial slope corresponding to $b \approx 0.394$ and $a \approx 0.301$. At longer times, $t \approx 100$, there appears to be a crossover to the logarithmic behavior associated with the strong disorder of figure 6.  

\begin{figure}[ht]
\includegraphics[width=7.5cm]{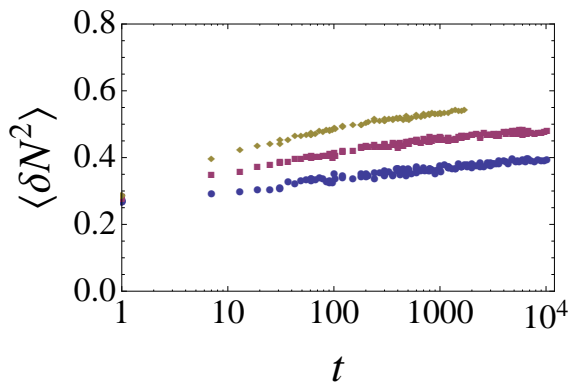}
\caption{\label{fig2} Comparison of variance for an $L=600$ lattice with three disorder strengths $\delta \kappa =0.5$ ($\diamond$), $0.75$ ($\sqcap$) and $1.0$ ($\circ$). }
\end{figure}

We first attempt to interpret this result by associating a dynamical exponent $z$ with the space and time scaling for the charge variance of disordered fermions. Taking the universal relation for the equilibrium charge variance at the IDFP ($b = \frac{1}{2} \log{2}$ in equation (\ref{spatial_fluct}), corresponding to the single boundary case) as the nominal spatial behavior, we write the temporal charge variance as follows:
\be
\label{dynamical_exponent}
\langle \delta N^2(t) \rangle = a + \frac{\log{2}}{2\pi^2} \log{t^{1/z}}
\ee
If systems $A$ and $B$ are connected at time $t=0$, $t^{1/z}$ represents the average distance to which fermion entanglements have propagated; $z=2$ would correspond to diffusive behavior, perhaps indicating diffusive propagation of the underlying entangled quanta. Using the fit to the data of figure 6, we find $z \approx 2.5$.  However there is a reasonable question about our association  single boundary RSP results for entanglement and noise (see the discussion in reference \cite{cardy_quench}). Comparing equations (\ref{FCS_CFT}) and  (\ref{spatial_fluct}) we see that in the clean case, dynamic fluctuations must be compared to the double boundary static case to yield dynamical exponent $z=1$. This feature is a consequence of the relevant conformal map used in the quench calculation. With disorder and the breakdown of conformal invariance, it is not clear that comparison with the double boundary case is appropriate.
 


Recently, Igloi and coworkers \cite{Igloi_disorder} have proposed a dynamical scaling at the IDFP to explain their results on entanglement entropy after a quench. Rather than replacing the spatial scale $l$ with $t^{1/z}$ as in equation (\ref{dynamical_exponent}), they proposed $l\sim (\log{t})^{1/\psi}$.  This scaling may be understood in terms of the Random Singlet Phase: bonds spanning a length $l$ have an exponentially small energy scale $E \sim \exp{(-l^\psi)}$, where $\psi = 1/2$.  Associating the quench time, $t$, with an inverse energy reproduces the proposed scaling. Combining $l\sim (\log{t})^{1/\psi}$ with equation (\ref{spatial_fluct}), the variance  is then expected to behave as:
\begin{equation}
\label{loglog}
\langle \delta N^2(t) \rangle = a + \frac{1}{\pi^2}\frac{b}{\psi}\log{\log{t}} 
\end{equation}

Looking at figure 6, the variance appears logarithmic over four decades; however extending our calculation out to double exponentially long times (as in reference \cite{Igloi_disorder}) and performing more disorder averaging over fewer points reveals a deviation from $\log{t}$ behavior. 

\begin{figure}[ht]
\includegraphics[width=7.5cm]{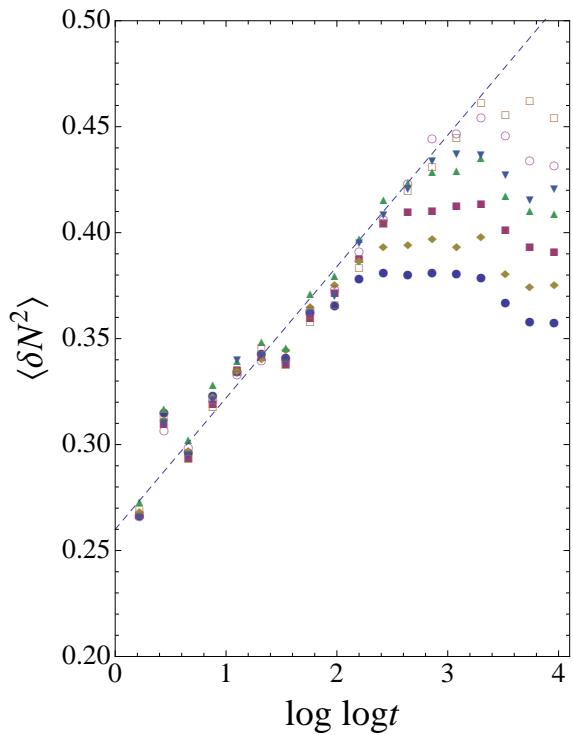}
 \caption{\label{fig2} (Color online) Variance as a function of  $\log\log{t}$, for a series of lattice sizes: (from bottom to top; number of realization shown in parentheses) $L = 48 \,(10000), 64 \,(10000), 96 \,(5000),128 \,(5000),192 \,(3000),$  $256 \,(3000), 600 \,(3000)$. In all calculations the disorder strength was $\delta \kappa =1.0$. Dashed line is a visual fit to the data envelope: $0.26+0.062\log\log{t}$.}
\end{figure}

Figure 8 shows the variance as a function of $\log\log{t}$ computed for several lattice sizes ranging from $L=48$ to $L=600$.  The envelope defined by these data sets is approximately linear and saturation of the variance appears to occur at doubly exponentially long times.  The saturation values appear to increase linearly in $\log{L}$. To extract the IDFP exponent, $\psi$, $b$ appearing in equation (\ref{loglog}) must be independently determined.  Direct analysis of the variance versus $\log{l}$ for a disordered half-chain of length $l$ yielded a value of $b \sim 0.365$ (equation (\ref{spatial_fluct}) and figure 5) close to the analytical result $b=\log{2}/2$.  Following \cite{Igloi_disorder} we have also determined $b$ from the saturations values of the variance implied by figure 8. 
\begin{figure}[ht]
\includegraphics[width=7.5cm]{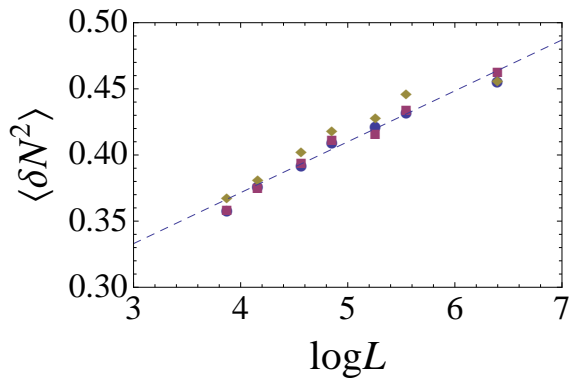}
 \caption{\label{fig2} (Color online) Saturation values of variance. Dependence of the variance at the three longest time values for the lattice sizes from figure 8 upon $\log{L}$.  Dashed line is least squares fit: $0.218 + 0.0385\log{L}$  }
\end{figure}
Figure 9 shows the variance for three largest time values plotted against $\log{L}$ for the system sizes, $L$, considered in figure 8. Fitting the data in figure 9 to equation (\ref{spatial_fluct}) yields a value of $b \approx 0.380$, a value close to that determined from figure 5.  Finally, $\psi$, from equation (\ref{loglog}) is determined to be $\psi \approx 0.621$.  Using the value of $b$ from figure 5, $\psi \approx 0.596$; using the value of $b$ from the exact IDFP ($b=\log{2}/2 \approx 0.347$), $\psi \approx 0.566$.  These values of $\psi$ appear to be close to the analytical value $\psi=1/2$ but systematically larger, a result also found in reference \cite{Igloi_disorder}.

We note that there appears to be systematic variations, independent of lattice size, in the variance for small times and for times larger than the saturation time, $t_{\rm sat}$.  For instance, there is a peak/dip structure at $t \sim 4.7 - 6.9$ and another at $t \sim 42 - 106$ that appears in all the lattice sizes studied.  There is also a systematic decrease in the variance at $\log\log{t} \sim 3.3-3.8$, beyond the saturation time for each lattice.  We speculate that the feature at $t \sim 4.7 - 6.9$ has to do with our choosing identical realizations of disorder for the $A$ and $B$ subsystems.  As discussed in section II, this was necessary to equalize the Fermi levels in both subsystems. The only reasonable explanation of the systematic feature at $t \sim 42 - 106$ is the presence of some ballistic fermions. Consistent with earlier computational work on 1-d localization, we believe there may be a substantial fraction of delocalized states with localization lengths in the range $\xi \sim 50 - 100$ lattice sites \cite{1d_disorder}.

\begin{figure}[ht]
\includegraphics[width=7.5cm]{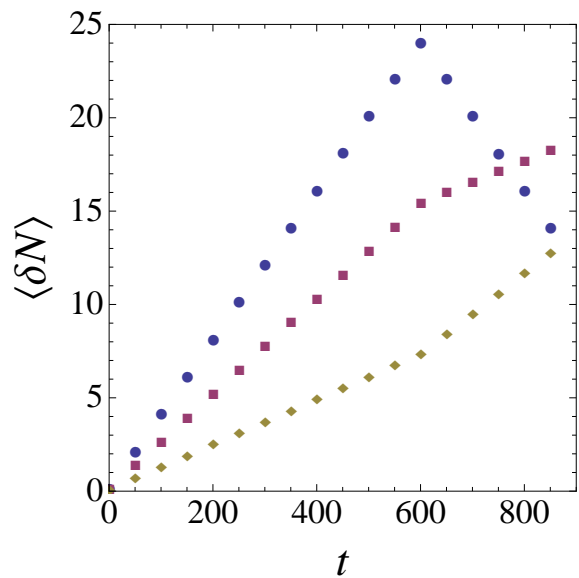}
 \caption{\label{fig2} (Color online) Fermions transferred to subsystem $A$ ($L=600$) as a function of time, $t$, for several weak link amplitudes, $w = \kappa, 0.5\kappa, 0.3\kappa$ (top to bottom.) Subsystems were prepared at $t=0$ with an initial bias $N_B - N_A = \Delta N = 24$.  The average number of fermions (first cumulant) is seen to increase linearly in time up to $t=t_b=L$, the ballistic time to encounter the boundary. For a transparent boundary $w=\kappa =1$, the first cumulant reaches a maximum $\langle \delta N \rangle = \Delta N$ at $t=t_b=L$ and then linearly returns to zero at $t = 2L$.  Because of partial reflection ($w \ne \kappa$), $\langle \delta N \rangle$ does not reach its maximum until several ballistic time periods, $t_b$, have passed. }
\end{figure}

\section{Finite bias voltage}

Since the fermions in the present study are completely localized, applying a bias voltage across the junction does not result in a constant current. Nonetheless it is interesting to determine the rate of charge transfer and the resulting shot noise in the 1-d strong insulating regime. Based upon our results for an unbiased junction, one might expect that the time dependent variance of localized fermions described by the IDFP would follow a  double logarithmic law whether or not there is a bias.  In contrast, we have found that applying a bias significantly changes the dynamic resulting in a time dependent variance that is proportional to $\log{t}$ rather than $\log \log{t}$.

\begin{figure}[ht]
\includegraphics[width=7.5cm]{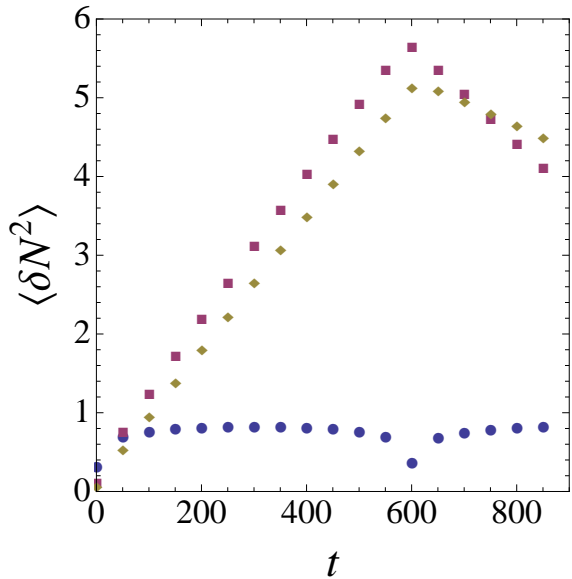}
 \caption{\label{fig2} (Color online)  time dependent variance for the same parameters as figure 10.  For $w=\kappa=1$, there is only the residual universal noise analogous to equation (\ref{FCS_CFT}) describing the unbiased case (compare with figure 4). For $w=0.5\kappa$ and $w=0.3\kappa$ the agreement of our computation with equation (\ref{noise_power}) is quite good. $\langle \delta N^2 \rangle$ should be linear in $t$ and achieve a maximum value of $|w^*|^2(1-|w^*|^2)\Delta N$.}
\end{figure}

Our numerical calculations employ open boundary conditions, as opposed to scattering boundary conditions, so a closed circuit bias voltage cannot be applied in the same way as analytical calculations on FCS. Instead a bias voltage is effected by preparing the A and B sides with different initial numbers of fermions and watching the system relax.  As a benchmark we first consider the disorder-free case.

For the results shown in Figure 10, two $L=600$ subsystems were prepared at time $t=0$ with different numbers of fermions: $N_A = L/2-\Delta N/2$ and $N_B = L/2+\Delta N/2$.  The initial fermion number difference $\Delta N$ was chosen to be $\Delta N = 24$ and the time evolution was studied for three different values of the weak link amplitude, $w$.  Figure 10 shows that the average number of fermions transmitted to subsystem $A$ (the first cumulant) increases linearly in time up to $t=t_b=L$, the ballistic time to encounter the boundary. For a transparent boundary $w=\kappa =1$, the first cumulant reaches a maximum $\langle \delta N \rangle = \Delta N$ at $t=t_b=L$ and then linearly returns to zero at $t = 2L$.  Because of partial reflection ($w \ne \kappa$), $\langle \delta N \rangle$ does not reach its maximum until several ballistic time periods, $t_b$, have passed.

To make connections with the Landauer formula for conductance \cite{Landauer} and the Lesovik formula for non-equilibrium noise power \cite{Lesovik}, we denote accumulated charge by $q \langle \delta N \rangle$. From figure 10 the initial current in the $w=\kappa$ case is $I = q \Delta N/L$.  Noting that the density of states for the half filled lattice is $L/2 \pi \kappa$, we arrive at the lattice I-V relation:
\begin{equation}
\label{IV}
I =|w^*|^2 \frac{q^2}{2\pi \kappa}V = g^*V
\end{equation}
where $|w^*|^2$ is the effective transmission coefficient at the Fermi level for the weak link and $g^*$ is the corresponding quantized conductance. Referring to figure 10, when $w=\kappa$ the effective transmission coefficient, $|w^*|^2 = 1$, as expected.  When $w=0.5\kappa$  and  $w=0.3\kappa$ the transmission coefficients are found to be $|w^*|^2=0.64$ and $|w^*|^2 = 0.3$, respectively. The detailed analytical relationship between lattice parameters and the transport parameters in equations (\ref{IV}) and (\ref{noise_power}) may be found in reference \cite{Schonhammer_num1}.

Adapting the Lesovik formula for noise power \cite{Lesovik} to our lattice calculation, we expect:
\begin{equation}
\label{noise_power}
\langle \delta N^2 \rangle =|w^*|^2(1-|w^*|^2)\frac{qV}{2\pi\kappa}t
\end{equation}
Figure 11 shows the time dependent variance for the same parameters as figure 10.  For $w=\kappa=1$, the shot noise predicted by equation (\ref{noise_power}) is zero and there is only the residual universal noise analogous to equation (\ref{FCS_CFT}) describing the unbiased case (compare with figure 4). For $w=0.5\kappa$ and $w=0.3\kappa$ the agreement of our computation with equation (\ref{noise_power}) is close. $\langle \delta N^2 \rangle$ should be linear in $t$ and achieve a maximum value of $|w^*|^2(1-|w^*|^2)\Delta N$ (converting voltage $V$ to equivalent $\Delta N$).  Using the values for $w^*$ found from figure 10 for the $w=0.5\kappa$ and $w=0.3\kappa$ cases, the maximum values of $\langle \delta N^2 \rangle$ are $5.53$ and $5.04$, respectively, in close agreement with figure 11. 

\begin{figure}[ht]
\includegraphics[width=7.5cm]{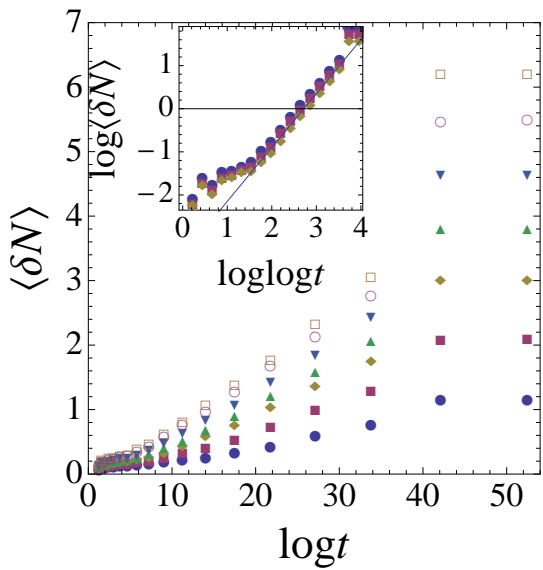}
 \caption{\label{fig2} First cumulant $\langle \delta N \rangle$ as a function of $\log{t}$ for a strongly disordered set of $L = 600$ lattices with (bottom to top) $\Delta N=4,8,12,16,20,24,28$. Each data set represents an average over $3000$ realizations of disorder ($\delta \kappa = 1$). Saturation appears at similar exponentially long times as in the unbiased case (figure 8). Inset: log-log plot of the data for the three largest $\Delta N$ to determine exponent $\psi$ in equation (\ref{cum1_bias}).  Solid line is a visual fit to the linear portion of the data with slope $\psi^{-1} = 1.25$.}
\end{figure}

We have computed the Full Counting Statistics of disordered lattices at finite bias using the same scheme as described before in section II.  Two identical realizations of an $L$-site disordered lattice are prepared with different numbers of fermions: $N_A = L/2-\Delta N/2$ and $N_B = L/2+\Delta N/2$ and connected at time $t=0$.  As discussed in section II, choosing {\sl different} realizations of disorder for $A$ and $B$ subsystems would  introduce a random bias voltage, $V_{\rm r}$, in effect.  To systematically study the effect of a bias voltage given the random fluctuations of $V_{\rm r}$, we would have had to bin our multiple realization results by the different values of $V$ requiring many more realizations to overcome the large statistical fluctuations inherent in a small bin size. 

\begin{figure}[ht]
\includegraphics[width=7.5cm]{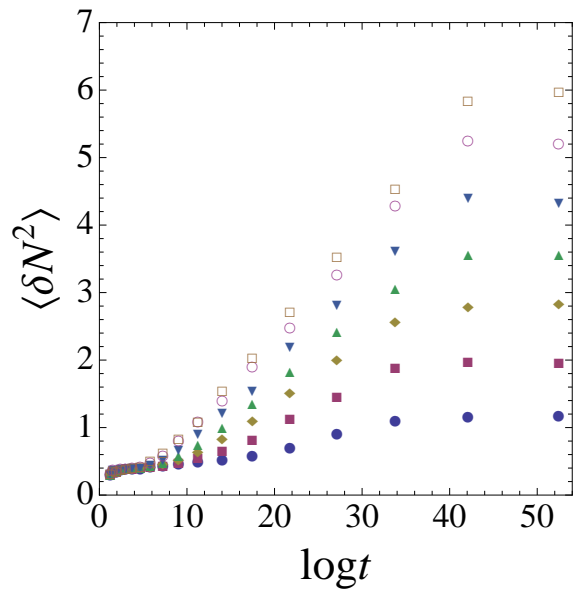}
 \caption{\label{fig2} (Color online) Variance as a function of $\log{t}$ for the same parameters as in figure 12. Data suggests a $\log\log{t}$ behavior for $\log{t} < \log{L}$ analogous to the unbiased case shown in figures 6 and 8. The variance appears to be proportional to $\log{t}$ for large times and saturate at exponentially long times consistent with the unbiased case (figure 8).  $\langle \delta N \rangle$ (figure 12) and $\langle \delta N^2 \rangle$ appear to saturate at the same value of $\Delta N/4$ suggesting that the saturation counting distribution is a poisson distribution.}
\end{figure}

Figures 12 and 13 show $\langle \delta N \rangle$ and $\langle \delta N^2 \rangle$  as a function of $\log{t}$ for a strongly disordered set of $L = 600$ lattices with several different $\Delta N$.  Saturation in both $\langle \delta N \rangle$ and $\langle \delta N^2 \rangle$ appears to occur at exponentially long times, similar in magnitude to the unbiased case.  Using IDFP scaling obtained from the unbiased case, we expect the saturation time, $t_{\rm sat} \sim \exp{L^\psi}$.  Therefore $\log{t_{\rm sat}} = 53$ or $\log{t_{\rm sat}} = 45$ using the two numerically determined values of the exponent $\psi = 0.621$ and $\psi=0.596$ from the unbiased variance data. These saturation times are in reasonable agreement with figures 12 and 13. Both $\langle \delta N \rangle$ and $\langle \delta N^2 \rangle$ appear to saturate at the value of $\Delta N/4$ suggesting that the saturation counting distribution is a Poisson distribution:
\begin{equation}
\log{\chi(\lambda,t\rightarrow \infty)} \approx \frac{\Delta N}{4}(e^{i\lambda}-1)
\end{equation}

Using  the scaling at the IDFP, one might expect that at a time $t$ after the junction is closed, fermions at a distance of $l \sim (\log{t})^{1/\psi}$ from the boundary are able to make a fluctuation bringing them to the boundary.  Then the number of charges accessible for tunneling would be: 
\begin{equation}
\label{cum1_bias}
\langle \delta N \rangle \sim  \Delta N \frac{l}{L} \sim \Delta N \frac{(\log{t})^{1/\psi}}{(\log{t_{\rm sat}})^{1/\psi}} 
\end{equation}
The curvature evident in figure 12 supports a power law interpretation.  The inset of figure 12 is a log-log plot of the data for the three largest voltages in figure 12, however a linear fit gives an estimate of $\psi = 0.8$, systematically larger than the earlier values determined by saturation data.

The variance $\langle \delta N^2 \rangle$, shown in figure 13, appears to have two regimes.  At large times running from approximately $\log{t}=8$ to $\log{t_{\rm sat}}$, $\langle \delta N^2 \rangle$ is proportional to $\log{t}$.  At short times, $\log{t} < \log{L}$,  is nearly constant and independent of voltage. The variance in this regime is very close to the unbiased case (see figure 6). Figure 14 shows the variance for an $L=300$ lattice with a comparable range of $\Delta N$; the time at which the crossover takes place appears to be smaller---possibly proportional to $\log{L}$.

Using the saturation value of the variance, $\Delta N/4$, and the linearity in $\log{t}$ we can formulate an empirical law for $\langle \delta N^2 \rangle$:
\begin{equation}
\label{variance_bias}
\langle \delta N^2 \rangle \sim \frac{1}{4} \Delta N \frac{\log{t}}{\log{t_{\rm sat}}} = L^{1-\psi}\frac{qV}{8\pi\kappa}
\end{equation}
In forming the first equality, we have ignored the short time ($\log{t} < \log{L}$) behavior.  In the second equality we have used the IDFP scaling for $t_{\rm sat}$ and expressed $\Delta N$ in terms of a voltage, $V$ using the density of states for the ordered lattice.  We note that the true density of states for a 1-d disordered lattice has a complicated singularity at the Fermi level and this second equality cannot be strictly true for $V\rightarrow 0$ \cite{Eggarter}.

\begin{figure}[ht]
\includegraphics[width=7.5cm]{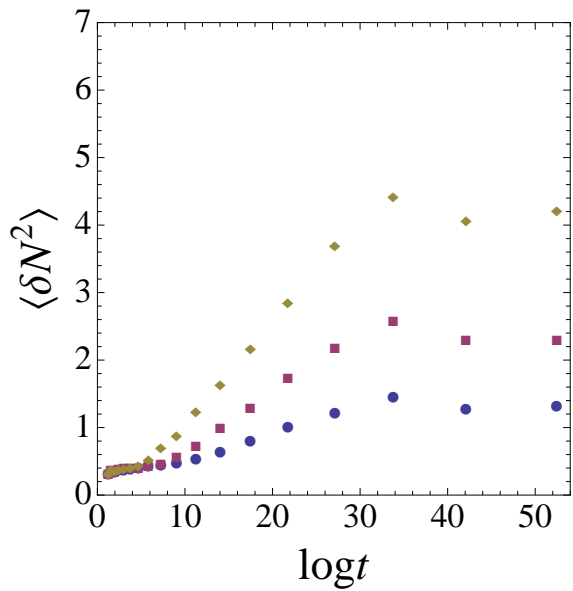}
 \caption{\label{fig2} (Color online) Variance as a function of $\log{t}$ for the same parameters as in figure 13 except $L=300$ and $\Delta N = 4,8,16$ have been changed to to effect similar voltages as figure 13. The crossover to linear behavior now occurs at smaller times, suggesting a crossover at $\log{t} \sim \log{L}$.  }
\end{figure}

\section{Conclusion}

In this manuscript we have made a computational study of charge fluctuations across a junction of two strongly disordered 1-dimensional fermion systems.  At zero bias, we have analyzed the variance as a function of time using a dynamical exponent, and using a recently proposed dynamical scaling at the IDFP \cite{Igloi_disorder}. By comparing to the known results for static charge fluctuations at the IDFP, we have computed the dynamical exponent and found $z \approx 2.5$.  However upon examining doubly exponential time scales, we come to the same conclusion as Igloi, et al concerning the dynamics of entanglement in reference \cite{Igloi_disorder}: a region of spatial scale $l$ relaxes with a time scale $t \sim \exp{l^\psi}$ correspondent with the exponentially small energy scales characteristic of the IDFP. This, in turn,  leads to the $\log \log {t}$ behavior of equation (\ref{loglog}).  We have also concluded that the saturation of the variance in a clean system appears to be well described by the universal CFT expression developed in references \cite{stephan_quench}. 

We have also studied the charge transfer and variance with an initial population imbalance, effecting an initial bias voltage. For clean systems our results closely follow Landauer formula for conductivity and Lesovik formula for noise, modified for our open boundary condition lattice geometry.  With disorder the fermions are localized and charge transfer may be thought of as activated by quantum fluctuations after the connection of the two subsystems. Similar to the zero bias case, charge and variance are seen to saturate at doubly exponential times reflecting the IDFP.  However, in contrast with the $\log \log {t}$ behavior for zero bias, charge and variance relax towards their saturation values as $(\log t)^{1/\psi}$ and $\log t$ respectively.  It is interesting to note, even with fully localized fermions, our results for FCS with an applied voltage involves the Fermi statistics implicit in the organization of the disordered fixed point.

GL wishes to thank Sasha Abanov for useful discussions and pointing out references \cite{Schonhammer_num1,Schonhammer_num2}.
This research was supported by the Department of Energy DE-FG02-08ER64623---Hofstra University Center for Condensed Matter and Research Corporation CC6535 (GL).

\end{document}